\documentclass[iop]{emulateapj}

\usepackage{amsmath}
\usepackage{natbib}
\usepackage{xcolor}
\usepackage[backref,breaklinks,colorlinks,citecolor=blue]{hyperref} 
\usepackage[all]{hypcap} 

\newcommand{\mps}{{\rm\,m\,s}^{-1}}
\newcommand{\mpss}{{\rm\,m\,s}^{-2}}

\newcommand{\trad}{\tau_{\rm{rad}}}

\usepackage{setspace}

\begin{document}

\title{Effects of latent heating on atmospheres of brown dwarfs and directly imaged planets }
\author{Xianyu Tan and Adam P. Showman}
\affil{Department of Planetary Sciences and Lunar and Planetary Laboratory, University of Arizona, 1629 University Boulevard, Tucson, AZ 85721, USA}
\email{xianyut@lpl.arizona.edu}


\begin{abstract}
Growing observations of brown dwarfs
have provided evidence for strong atmospheric circulation on these objects. Directly imaged planets share similar observations, and  can be viewed as low-gravity versions of brown dwarfs. Vigorous condensate cycles of chemical species in their atmospheres are inferred by observations and theoretical studies, and  latent heating associated with condensation is expected to be important in shaping atmospheric circulation and influencing  cloud patchiness. We present a qualitative description of the mechanisms by which condensational latent heating influence the circulation, and then illustrate them using an idealized general circulation model that includes a condensation cycle of silicates with latent heating and molecular weight effect due to rainout of condensate. Simulations with conditions appropriate for typical T dwarfs exhibit the development of localized storms and east-west jets. The storms are spatially inhomogeneous, evolving on timescale of hours to days and extending vertically from the condensation level to the tropopause. The fractional area of the brown dwarf covered by active storms is small. Based on a simple analytic model, we quantitatively explain the area fraction of moist plumes, and show its dependence on radiative timescale and convective available potential energy. 
We predict that, if latent heating dominates cloud formation processes, the fractional coverage area by clouds decreases as the spectral type goes through the L/T transition from high to lower effective temperature. This is a natural consequence of the variation of radiative timescale and convective available potential energy with spectral type.
\end{abstract}

\section{INTRODUCTION}
Observations of brown dwarfs (BDs) have shown increasing evidence of a vigorous  circulation in their atmospheres \citep{showman&kaspi2013}. This evidence includes near-infrared brightness variability \citep{artigau2009,radigan2012,apai2013,buenzli2014,radigan2014,wilson2014,buenzli2015,metchev2015,yang2015, yang2016, cushing2016}, chemical disequilibrium \citep{fegley1996, saumon2006, saumon2007, hubeny2007, stephens2009,visscher2011,zahnle2014} and surface patchiness \citep{crossfield2014}.  Cloud disruption has been proposed to help explain properties of the L/T transition \citep{ackerman2001, burgasser2002,marley2010}, and such patchiness is also likely responsible for  the near-infrared brightness variability \citep{marley2015}.  Nevertheless, the mechanism responsible for cloud disruption is yet unclear. Atmospheric circulation is expected to play a crucial role in controlling cloud coverage fraction, but the details remain poorly understood. 

A handful of directly imaged extrasolar giant planets (EGPs)  exhibit similarities with BDs: the near-infrared colors, inference of dust and clouds, chemical disequilibrium in their atmospheres and fast spin \citep{hinz2010,barman2011a,barman2011b,marley2012,oppenheimer2013,ingraham2014,skemer2014,snellen2014, macintosh2015, wagner2016}. Near-IR brightness variability has also recently been observed on directly imaged EGPs \citep{biller2015, zhou2016}. From a meteorological point of view,  the directly imaged EGPs resemble  low-gravity versions of BDs, for which their atmospheric dynamical regime is characterized by fast rotation, vigorous convection and negligible external heating.

Motivated by the observations, several studies have been conducted to explore the atmospheric dynamics of  ultra cool objects  (compared to stars). Local two-dimensional hydrodynamics simulations by \cite{freytag2010} showed that interactions between the convective interior and the stratified layer can generate  gravity waves that propagate upward, and the breaking of these waves causes vertical mixing that leads to small-scale cloud patchiness.  \cite{showman&kaspi2013} presented the first global model of brown dwarf dynamics for the convective interior, and showed that large-scale convection is dominated by the fast rotation. Using an analytic theory, they proposed that atmospheric circulation can be driven by atmospheric waves in the stably stratified upper atmosphere.  Using a two-layer shallow-water model,  \cite{zhang&showman2014} showed that weak radiative dissipation and strong forcing  favor large-scale zonal jets for brown dwarfs, whereas strong dissipation and weak forcing favor transient eddies and quasi-isotropic turbulence. 
Despite these studies, no global model  that includes condensate cycles and clouds has yet been published for brown dwarfs.  Clouds play a significant role in sculpting the temperature structure, spectra and brightness variations of brown dwarfs (see recent reviews of \citealp{marley2015} and \citealp{helling2014}).  There is a pressing need to couple condensation cycles and clouds to global models  to study how the circulation controls global cloud patchiness, and in turn how the condense cycle affects the circulation.

In this paper, we propose  the importance of  latent heating on the atmospheric circulation and cloud patchiness of brown dwarfs by using an idealized general circulation model that includes a condensation cycle of silicate vapor.  Latent heating is of paramount importance in Earth's atmosphere \citep{emanuel1994}. For giant planets in our solar system whose atmospheres are likely analogous to brown dwarfs',   a long history of studies has shown the importance of latent heating in driving their atmospheric circulation (\citealp{barcilon1970, gierasch1976, gierasch2000, ingersoll2000, lian&showman2010}). \cite{lian&showman2010} demonstrated that large-scale latent heating from condensation of water can drive patterns of zonal (east-west) jet streams that resemble those on all four giant planets of the solar system: numerous zonal jets off the equator and a strong prograde equatorial jet on Jupiter and Saturn, and a three-jet pattern including retrograde equatorial flow and high-latitude prograde flow on Uranus and Neptune. Such models also exhibit episodic storms that qualitatively resemble those observed on Jupiter and Saturn.  For brown dwarfs, the evidence for patchy clouds in controlling brightness variability and the L/T transition itself also suggests a strong role for an active condensate cycle, and latent heating may be similarly important for atmospheric circulation of BDs and directly imaged EGPs. Because temperature perturbations associated with (dry) convection at condensable pressure levels are generally small, the latent heating that accompanies the condensation of relevant chemical species  can dominate the buoyancy in the layers where condensation occurs. 

The main point of this paper is to illustrate how latent heating modifies a circulation and influences cloud patchiness in the simplest possible context, so we intentionally exclude clouds, radiative transfer and detailed microphysics to allow a simpler environment in which to clarify the dynamical processes that are at play. Cloud microphysics processes are highly complex  \citep{rossow1978}, and significant prior work on the cloud microphysics issue (see a review by \citealp{helling2014}), as well as parameterized cloud models \citep{allard2001,ackerman2001, tsuji2002, cooper2003, barman2011a} has been done for ultra cool atmospheres.  We are well aware of the important feedback of clouds to  atmospheres, and will leave it for future efforts. Also, to resemble the vigorous convection and the dynamics in radiative-convective boundary caused by convective perturbation, one needs a model that can properly treat both the convective interior and the overlying stably stratified layer. Therefore, we do not expect our current simulations to resemble the true atmospheres of brown dwarfs and directly imaged EGPs.

The paper is organized as follows. We start out  in Section \ref{mechanism} by describing several important effects of latent heating  on the atmosphere; in Section \ref{model}, we briefly introduce our idealized model that is used to illustrate the mechanisms described in Section \ref{mechanism}; in Section \ref{results}, we show result of our simulations; finally in Section \ref{discussion}, we discuss our results and implications for observations, and draw conclusions.

\section{Effects of latent heating on atmospheres}
\label{mechanism}
\subsection{Conditional Instability}
\label{instability}
Most atmospheres of planets and ultra cool brown dwarfs have constituents that can condense. Due to atmospheric motion and diabatic heating/cooling, air parcels containing condensable species can undergo change of temperature and pressure, leading to condensation. The latent heating/cooling due to condensation/evaporation has important effects on the stability of atmospheres, which we summarize here; a more detailed discussion can be found, e.g., in Chapter 7 of \citet{salby2012}. For simplicity, we begin our discussion assuming the molecular weight is constant but return to this issue in a later subsection. 

It is well known that a rapidly ascending or descending dry air parcel follows a dry adiabatic lapse rate  
\begin{equation}
\frac{d\ln T}{d\ln p} = \frac{R}{c_p},
\end{equation}
where $T$ is temperature, $p$ is pressure, $R$ is specific gas constant and $c_p$ is specific heat capacity of dry atmosphere. Similarly for a saturated air parcel mixed with condensable and non-condensable gases, it follows a moist adiabatic lapse rate\footnote{In deriving this formula, the Clausius-Clapeyron equation for the saturation vapor pressure of the condensable species was used, assuming an ideal gas equation of state and that the condensate density is much greater than the gas density. This formula is applicable for full range of $\xi$, not limited to assumption of small mixing ratio of condensable gas.}:
\begin{equation}
\frac{d\ln T}{d\ln p} = \frac{R_u+\frac{L_m\xi}{T}}{c_p+\frac{L_m^2\xi}{R_uT^2}},
\end{equation}
where $L_m$ is the latent heat per mole, $c_p$ is the specific heat capacity per mole for the mixture, $R_u$ is the universal gas constant, and $\xi = p_{cond}/p_d$ is the molar mixing ratio of condensable gas over dry gas. Under normal conditions of most atmospheres, the dry adiabat is larger than the moist adiabat as long as $L_m > c_pT$. In the presence of two adiabats, the atmosphere can have different stability criteria. If the atmospheric lapse rate $\frac{d\ln T}{d\ln p}$ is larger than the dry adiabatic lapse rate $R/c_p$, the atmosphere will be absolutely unstable; if the lapse rate is smaller than the moist adiabatic lapse rate, the atmosphere will be absolutely stable; if the lapse rate is in between the dry and moist adiabatic lapse rate, the atmosphere is stable against dry convection but unstable to moist convection, which is referred to as  \emph{conditional instability}. Examples include the tropospheres of the Earth, Titan, and probably Jupiter and Saturn.

\begin{figure*}      
\epsscale{1.}      
\plotone{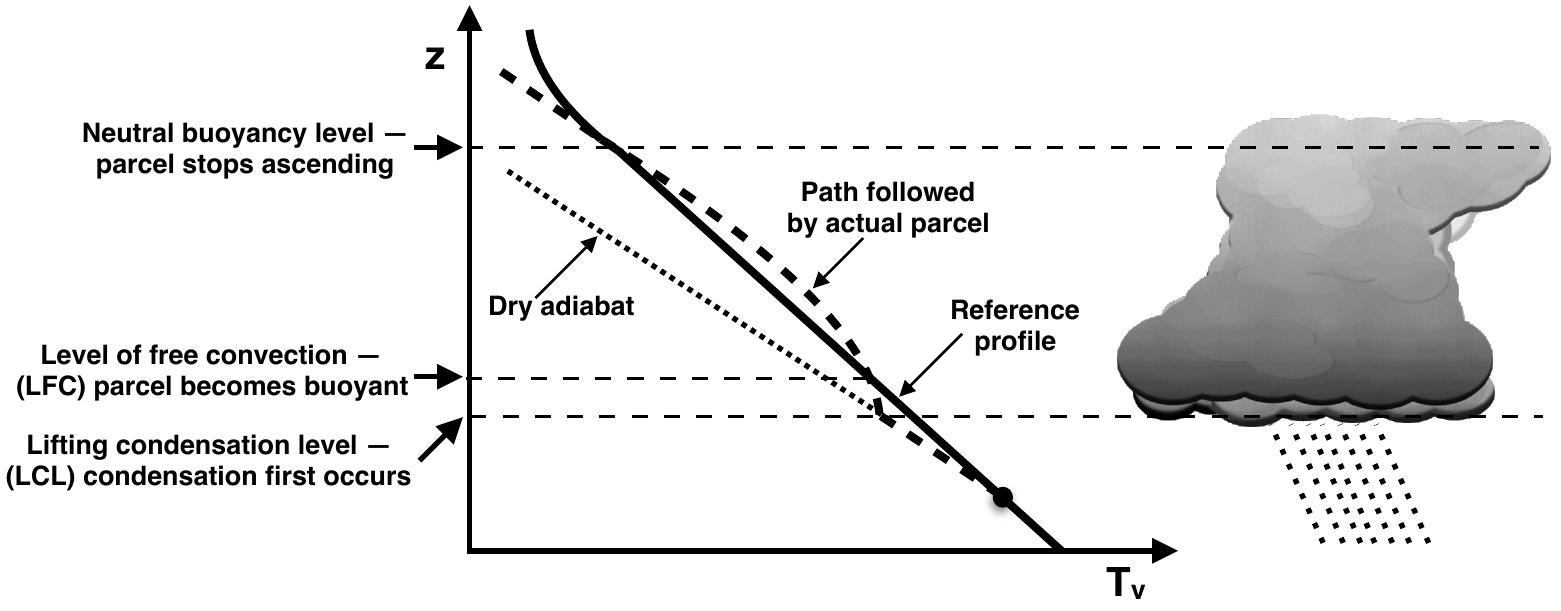}
\caption{A schematic plot showing how an air parcel behaves in a conditionally unstable atmosphere. The solid line is the background reference temperature profile,  the dashed line is the path followed by an actual air parcel initiated below the lifting condensation level, and the dotted line indicates the dry adiabatic path if the air parcel is dry. Horizontal axis is temperature (or, if molecular weight effects are important, the virtual temperature), and the vertical axis is height within the atmosphere.}
\label{schematic}
\end{figure*} 

How does an air parcel  behave in a conditional unstable and unsaturated atmosphere? This is schematically illustrated in Figure \ref{schematic}: initially starting from an arbitrary level below the condensation level,  the ascending air parcel follows a dry adiabat until its relative humidity reaches 100\%, and reaches the lifting condensation level (LCL). Afterward it will follow a moist adiabat, and then at some point it will reach the level of free convection (LFC) where it has a lower density than the environment and becomes positively buoyant. The air parcel then can freely convect to the top of a cumulus storm where its buoyancy diminishes and it stops ascending. In reality,  because the atmospheric lapse rate may be stable to dry convection, some external lifting mechanism is needed for the air parcel to reach  the LFC, and this is why storms do not occur everywhere and in every moment in Earth's tropics even though the atmosphere is conditionally unstable. The amount of energy required to reach the LFC is referred to as convective inhibition (CIN). To initiate moist convection, either strong initial diabatic heating (e.g., in the case of Earth, heating of the surface by sunlight) or  kinetic energy (e.g., forced lifting by atmospheric waves or other large-scale motion) is needed for the parcel to overcome the CIN. The CIN can act to limit the frequency of  moist convection and preserve large convective available potential energy, which can be essential for the development of deep moist convection.

One necessary condition for conditional instability is the  stratification to dry convection in the troposphere. One mechanism to produce  the tropospheric stratification is by latent heating and moist convection. As illustrated in Figure \ref{schematic}, the rising air that follows a moist adiabat  in storms carries a higher entropy than where it is initiated. Mass continuity implies that the high-entropy surrounding atmosphere at the top of the storm must subside. During the subsidence, air continues to lose its entropy to space via IR radiation. Air closer to the ground has been subsiding for longer -- and thus exhibits lower entropy -- than air aloft.  As a result, the entropy increases with height in the background temperature profile, which implies that the background temperature profile is stable to dry convection. On the other hand, the background temperature exhibits lower temperature than the moist adiabat, allowing moist instability.  For giant planets, the tropospheric stratification  has been inferred from observations for Jupiter  \citep{flasar1986, magalhaes2002, reuter2007}, and it has been demonstrated by numerical simulations that the stratification in Jupiter can result from latent heating of water condensation \citep{nakajima2000,sugiyama2014}. 

\subsection{Moist Convection on Controlling the Area Fraction of Moist Plumes}
Vertical velocity within the moist convecting plume is much larger than the surrounding subsidence flow, as observed in Earth's atmosphere. The fast upwelling velocity is driven by the large convective available potential energy (CAPE) in deep moist convection. CAPE is the amount of potential energy per unit mass available for the convection of a particular air parcel, and is essentially an integration of buoyancy with respect to height during the lifting of the parcel (e.g., \citealp{emanuel1994}, Chapter 6).  In contrast to the buoyant ascending convective plumes, the subsiding air is stratified and not convecting, but it can gradually subside as described in Section \ref{instability}. Because the radiative cooling time scale is generally long,  the subsidence is generally slow. The asymmetry of the upwelling and downwelling vertical velocity results in a small area fraction of moist ascending plumes by the requirement of mass continuity. The area covered by clouds  may not closely follow the area of moist plumes because cloud particles will be spread by the wind field near the cloud top, but this argument for moist plume area fraction can qualitatively explain the origin of patchiness of cumulus clouds \citep{lunine1987}. For brown dwarfs, patchy clouds deduced from near-IR brightness variability may  qualitatively be explained by this mechanism.

\subsection{Molecular Weight Effect}
\label{molecular}
In atmospheres of gaseous giant planets and BDs, condensable species generally have a much higher molecular weight than the dominant dry constituent $\rm{H}_2$. Rainout of these condensates can decrease the density of air, and this effect can play an important role in atmospheric  thermal structure and dynamics. For example,  \cite{guillot1995} presented the idea that moist convection in giant planets may be inhibited due to molecular weight effect if the mixing ratio of water is substantially higher than solar abundance; \cite{Li2015} proposed that Saturn's  20-to-30-year quasi-periodic planetary-scale storm is related to the molecular weight effect of water. 
In BDs and directly imaged EGPs, a local thin stably stratified layer associated with molecular weight gradients may exist right above the condensation level,  due to the fact that the subsiding environmental air has experienced rainout, is thus relatively dry and has lower molecular mass. Whereas the air below the condensation level contains significant condensable vapor and has higher molecular mass. Therefore, a sharp gradient in molecular mass naturally exists near the condensation level, where molecular mass decreases with increasing altitude. This produces  stratification to both dry and moist convection, contributing to CIN. Large CIN can suppress moist convection within (and above) the stable layer, leaving the atmosphere gradually cooling off towards its radiative equilibrium by thermal radiation. The radiative equilibrium temperature profile in the troposphere (for instance, on Earth and Jupiter) usually has a temperature lapse rate $d\ln T / d\ln p$ larger than the moist adiabatic lapse rate. Thus, the existence of CIN can help contribute to the accumulation of CAPE.
This phenomenon has been shown by simulation on Jupiter as well  \citep{nakajima2000}. By contribution to the accumulation of CAPE,  the stratification from molecular weight effect also help to control the moist plume fraction via the mechanism discussed above.

\subsection{Large-scale Latent Heating on Atmospheric Dynamics}
\label{latentheating}
Moist convection provides a source of small-scale eddies, which can grow into large-scale eddies via an inverse energy cascade. The interactions among these eddies and the mean flow in a rapidly rotating sphere can produce zonally banded structure and vortices  (\citealt{lian&showman2010}, also see a review by \citealt{vasavada2005} for Jovian atmospheric dynamics). The latent heating can interact with the dynamics in many ways, and may produce organized clouds that can lead to cloud radiative feedback to the dynamics. The temperature perturbations by latent heating on isobaric surface can generate a wealth of waves that propagate upward to the stratosphere, driving circulation by the dissipation and breaking of these waves \citep{showman&kaspi2013}.

\section{MODEL}
\label{model}

Here we summarize the key aspects of our model; for detailed implementation see \cite{lian&showman2010}.  We solve the three-dimensional hydrostatic primitive equations using an atmospheric general circulation model (GCM), the MITgcm (\citealp{adcroft2004}, see also mitgcm.org). The horizontal momentum, hydrostatic equilibrium, continuity, thermodynamic energy and tracer equations in pressure coordinates are, respectively,
\begin{equation}
\frac{d\mathbf{v}}{dt} + f\hat{k} \times \mathbf{v} + \nabla_p \Phi = 0, 
\label{eq.momentum}
\end{equation}
\begin{equation}
\frac{\partial \Phi}{\partial p} = -\frac{1}{\rho}, 
\label{eq.hydro}
\end{equation}
\begin{equation}
\nabla_p \cdot \mathbf{v} + \frac{\partial\omega}{\partial p} = 0,
\label{eq.cont}
\end{equation}
\begin{equation}
\frac{d\theta}{dt} = -\frac{\theta-\theta_{\rm{ref}}}{\trad}  + \frac{L\theta}{c_p T}(\delta \frac{q-q_s}{\tau_{\rm{cond}}}),
\label{thermal}
\end{equation}
\begin{equation}
\frac{dq}{dt} = -\delta\frac{q-q_s}{\tau_{\rm{cond}}}+Q_{\rm{deep}},
\label{tracer}
\end{equation}
where $\mathbf{v}$ is the horizontal velocity vector on isobars, $\omega=dp/dt$ is the vertical velocity in pressure coordinates, $f=2\Omega\sin \phi$ is the Coriolis parameter (here $\phi$ is latitude and $\Omega$ is the planetary rotation rate), $\Phi$ is geopotential, $\hat{k}$ is the local unit vector in the vertical direction, $\rho$ is density, $\nabla_p$ is the horizontal gradient in pressure coordinate, $d/dt=\partial/\partial t + \mathbf{v}\cdot\nabla_p + \omega\partial/\partial p$ is the material derivative, $\theta = T(\frac{p_0}{p})^{R/c_p}$ is the potential temperature,   $p_0 = 1$ bar is a reference pressure, $\theta_{\rm{ref}}$ is the equilibrium potential temperature profile, $\trad$ is the radiative timescale, and $L$ is latent heat per mass for condensate. The ideal gas law is assumed for the equation of state for the atmosphere.

The hydrostatic assumption used in the standard primitive equations solved in our model is a good approximation for large-scale flows in stratified atmospheres with large ratio of horizontal scale to vertical scale (e.g., see \citealp{vallis2006}, Chapter 2). In the atmospheres of brown dwarfs, the expected horizontal length scale of large-scale dynamics is $10^6 - 10^7$ meters, while the  pressure scale height is $10^3 - 10^4$ meters. The aspect ratio of the atmosphere is on the order  $10^2 - 10^3$, which is sufficient for the hydrostatic approximation to hold.

\begin{figure}      
\epsscale{1.}      
\plotone{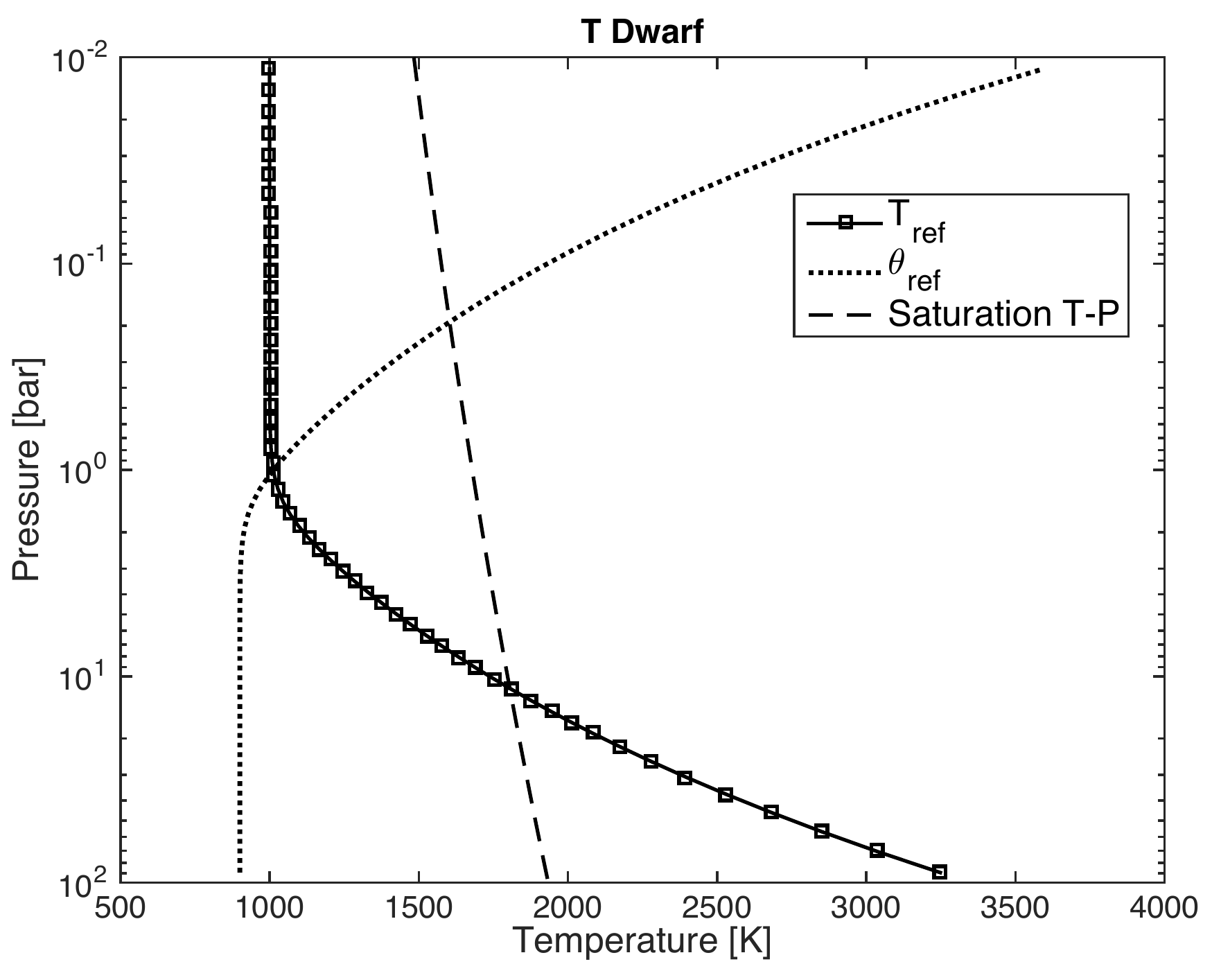}
\caption{Equilibrium T-P profile $T_{\rm{ref}}$ in the model (solid line) with squares representing the model pressure levels, and the saturation  T-P profile (dashed line) of enstatite ($\rm{MgSiO_3}$) for  a typical T dwarf temperature regime assuming $1\times$ solar abundance. The dotted line is the corresponding potential temperature $\theta_{\rm{ref}}$ profile.}
\label{f_condensate}
\end{figure}

The tracer $q$ is mass mixing ratio of condensable vapor to dry air, and  $q_s$ is the local saturation vapor mass mixing ratio that is determined by saturation pressure function for specific condensable species.    The ``on-off switch'' function $\delta$ controls the condensation: when $q>q_s$ then $\delta=1$ and vapor condenses over a characteristic  timescale $\tau_{\rm{cond}}$ which is generally taken as $10^3$ sec, representative of a typical convective time; when $q \leq q_s$ then $\delta = 0$. Latent heating is immediately applied in the thermodynamic equation (Equation [\ref{thermal}]) once condensation occurs. For simplicity, we include only one tracer, and choose  enstatite vapor ($\rm{MgSiO_3}$) to represent silicate vapor in our brown dwarf models. Silicates are one of the most abundant condensates in the atmospheres of L/T dwarfs, and their condensation levels are closer to the photospheres than another dominant condensates -- iron, and so silicates could have more influences on the photospheres of L/T dwarfs than iron. The saturation pressure function for $\rm{MgSiO_3}$ is adopted from \cite{ackerman2001}. Alternative saturation pressure functions for silicates are available, for example, see \cite{visscher2010}. Our study does not aim at precisely determining where condensation occurs but rather to explore dynamics driven by latent heating given a plausible condensation curve for a representative condensing species. The latent heat of silicates are similar in \cite{ackerman2001} and \cite{visscher2010},  therefore, the detailed choice of the saturation T-P profile is not essential here. The saturation pressure function reads 
\begin{equation}
e_{\rm{s}} = \exp(25.37 - \frac{58663~\rm{K}}{T}) \quad \rm{bar},
\end{equation}
which is shown by the dashed line in Figure \ref{f_condensate} assuming  solar abundance for the mixing ratio of silicates. Here we assume  that condensate will rain out immediately. The influence of rainout of condensable vapor on air density is properly included in the hydrostatic equilibrium equation where the density is affected by mean molecular weight. 
The replenishment term $Q_{\rm{deep}}$ crudely parameterizes evaporating precipitation and condensable species mixed upward from the deeper atmosphere. It takes the form $Q_{\rm{deep}} = (q_{\rm{deep}}-q)/\tau_{\rm{rep}}$, where $q_{\rm{deep}}$ is a specified abundance of condensable species in the deep atmosphere and $\tau_{\rm{rep}}$ is the replenish timescale which is typically taken $10^3$ sec. Both $\tau_{\rm{cond}}$ and $\tau_{\rm{rep}}$ are chosen to be short compared to dynamical timescales, and in this limit the dynamics should be independent of the two timescales. The $Q_{\rm{deep}}$ term is applied only at levels deeper than the condensation level. 

The radiation effects of the system are simplified by using the Newtonian cooling scheme (Equation [\ref{thermal}]).   For simplicity, the radiative timescale $\trad$ is taken to be constant through the atmosphere.  In our application, the radiative equilibrium potential temperature $\theta_{\rm{ref}}$ is assumed spherically symmetric, and is characterized by two regimes, a nearly adiabatic deeper region and an isothermal upper region as $\theta_{\rm{ref}}(p)=[\theta_{\rm{adi}}^n(p)+\theta_{\rm{iso}}^n(p)]^{1/n}$,  where $\theta_{\rm{adi}}$ represents  the potential temperature of the nearly adiabatic lower layer, $\theta_{\rm{iso}}$ represents that of the isothermal upper layer, and $n$ is a smoothing parameter that we here set to 15. 
The equilibrium temperature profile is intended to crudely mimic the results from one-dimensional radiative-convection models, where the profile of the upper atmosphere approaches nearly isothermal  and smoothly transitions to an adiabatic profile in the lower atmosphere (e.g., \citealp{marley2002,burrows2006,morley2014}). Our equilibrium temperature structure is based on a gray radiative-convective calculation using the Rosseland-mean opacity table from \cite{freedman2014}, and the radiative-convective boundary from our calculation is in good agreement with  models using realistic opacities (e.g., \citealp{tsuji2002}).  The deep thermal structure is generally slightly unstable rather than strictly neutral to allow dry convective motions. We parameterize the temperature structure of the adiabatic deep region by 
\begin{equation}
\theta_{\rm{adi}}(p) = \theta_0 + \delta \theta \log \frac{p}{p_{bot}}
\label{eq.theta}
\end{equation}
where $\theta_0$ and $\delta \theta$ are constants and $ p_{bot}$ is the bottom pressure of the simulation domain. $\delta \theta$ is typically taken as 1 K for 1 times solar cases, which is qualitatively consistent with the argument in mixing length theory, but small enough to not affect the dynamics above the condensation level.
 Figure \ref{f_condensate} shows the equilibrium temperature  and saturation vapor T-P profile for a typical T dwarf temperature regime, with silicates' condensation T-P curve. The dotted line is the corresponding potential temperature $\theta_{\rm{ref}}$ profile, in which the adiabatic layer is characterized by nearly constant $\theta$ and the isothermal layer has an increasing $\theta$ with increasing altitude.

Real moist convection involves the formation of cumulus clouds and thunderstorms on a length scale much smaller than that can be resolved by most general circulation models. There has been a long history of development for schemes that parameterize the effects of sub-grid-scale cumulus convection on large-scale flows resolved by global models (for a review see, e.g., \citealt{emanuel1993}). These schemes are often complex, with concepts and parameterizations  constrained by Earth's atmosphere. It is not yet clear how relevant the specific parameterization of these schemes is to  atmospheres of brown dwarfs and giant planets, so we do not include a moist convection sub-grid-scale scheme in our current model. As stated in \citet{lian&showman2010}, it is useful to first ascertain the effects of \emph{large-scale} latent heating associated with the hydrostatic interactions of storms with the surroundings, as we pursue here. 

We include a weak linear damping of velocities similar to that of \cite{liu2013} at pressure larger than 50 bars to mimic the reduction of winds due to the Lorentz force and Ohmic dissipation at great depths where magnetic coupling may be important. This drag is deep and weak enough (with drag timescale of 100 days at the bottom)  not to affect the dynamics above condensation level.  

We solve the equations of our global model on a sphere using the cube-sphere coordinate system \citep{adcroft2004,showman2009}.  For most of simulations, we assume a Jupiter radius, a five hour rotation period and 500 $\mpss$  surface gravity. The resolution in our nominal simulations is C128, which is equivalent to $0.7^{\circ}$ per grid longitudinally and latitudinally (i.e., an approximate resolution of $512\times256$ in longitude and latitude). The pressure domain in our model is from $0.01$ bar to $100$ bars, and it is divided into 55 layers with finer resolution on condensation layers as shown in Figure \ref{f_condensate}. The horizontal and vertical resolution is adequate to resolve the Rossby deformation radius which is the typical length scale of eddies expected on brown dwarfs, and the vapor partial pressure scale height above the condensation level, respectively. We do not include an explicit viscosity in our simulations, but a fourth-order Shapiro filter is added to the time derivative of $\mathbf{v}$ and $\theta$ to maintain numerical stability.

\begin{figure*}      
\epsscale{1.}      
\plotone{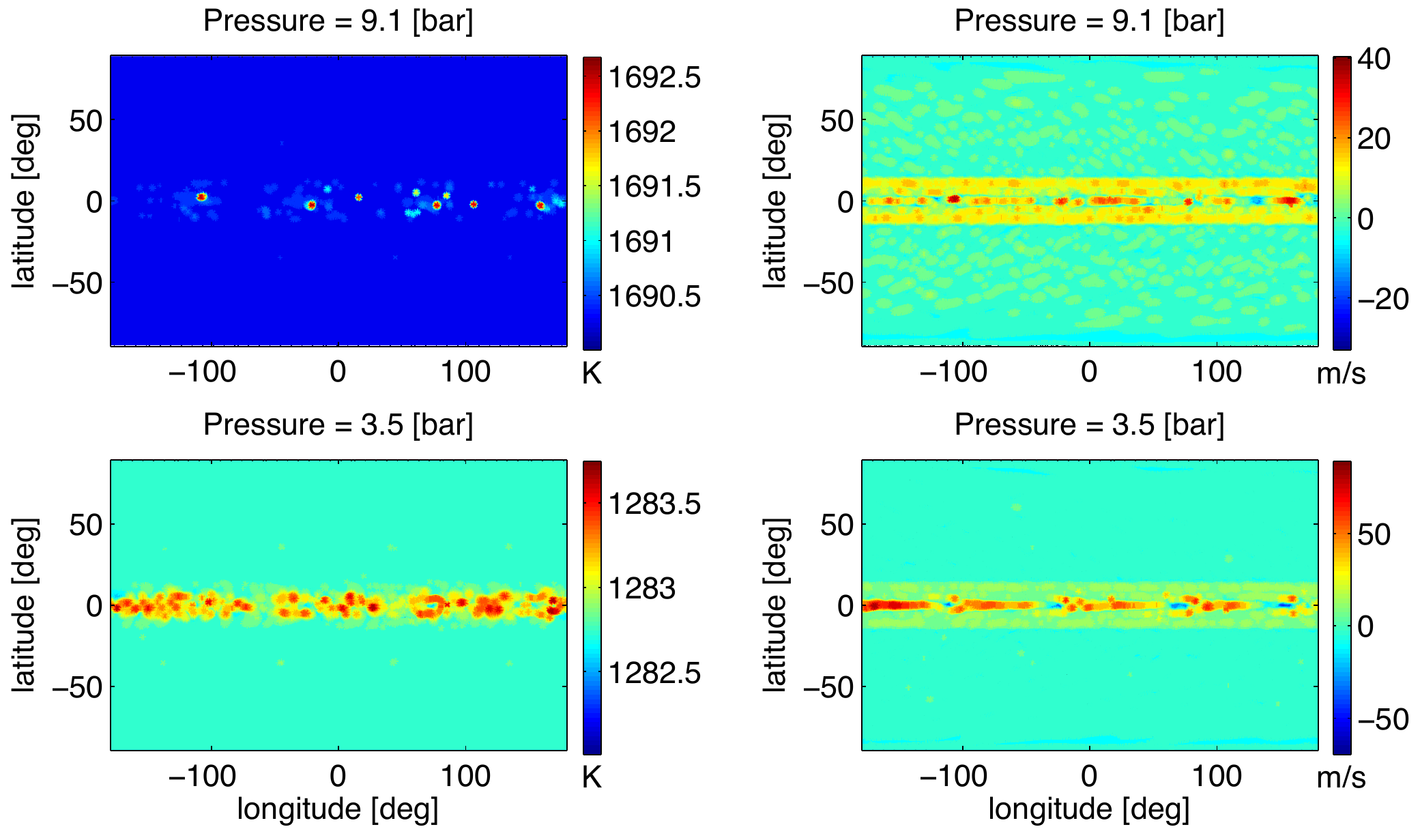}
\caption{\emph{Upper row}: snapshot of horizontal map of temperature (left, K) and zonal (that is, east-west) velocity (right, $\mps$) as a function of longitude and latitude at pressure level of 9.1 bars at about 1736 Earth days simulation time in an atmospheric circulation model for conditions appropriate to a typical brown dwarf. \emph{Lower row}: same as the upper row but at pressure level of 3.5 bar, near the tropopause.}
\label{f_bd15-1}
\end{figure*} 

\section{RESULTS}
\label{results}

\subsection{A Typical T Dwarf}
\label{tdwarf}
We begin by describing in detail a specific representative case for a typical T dwarf with a radiative timescale $\trad$ of $10^6$ sec,  solar metallicity which is typical for field brown dwarfs (e.g., \citealp{leggett2010}),  and other parameters described in Section \ref{model} (see also Figure \ref{f_condensate}). The spin-up time is about 1500 days for models with $\trad=10^6$  and $10^7$ sec and about 1000 days for model with $\trad=10^5$ sec. Figure \ref{f_bd15-1} shows a snapshot of a horizontal map of temperature and zonal (east-west) velocity at 1736 Earth days simulation time at 9.1 bars (upper row) near the condensation level. The simulation  reached a statistically equilibrium state, where latent heating from the condensate cycle is statistically balanced by radiative cooling, and the upward transport of condensable vapor is balanced by rain out in storms. On the temperature map (left panel), the local red regions are storms with warm upwelling moist plumes, and they evolve on a timescale of hours to (Earth) days. The upper right panel  in Figure \ref{f_bd15-1} shows the zonal wind map at the same pressure level, with yellow and red colors representing eastward velocity.  Three eastward jets form near the equator, with  maximum wind speed of about 40 $\mps$. The jets are located where storms are generated, suggesting that jets are pumped by momentum transport associated with the storms. No jets form at mid-to-high latitudes, but velocity residuals manifest there, which are Rossby waves propagating northward and southward from the storm regions. 

The upwelling vertical motions are strongly suppressed near the stably stratified isothermal layer at $p<2$ bars, causing large horizontal velocity divergence; as a result, the wider spreading of the upwelling hot air produces the larger temperature perturbation patterns near the tropopause (lower-left panel in Figure \ref{f_bd15-1}), similar to simulations for the Jupiter model \citep{lian&showman2010}. 
Because the ascending air inside storms extend vertically from the storm base to the top near the tropopause, the locations of warm regions at 3.5 bars are generally correlated to those at about 9 bars.   The horizontal zonal velocity map at the tropopause exhibits the similar multiple-jet configuration as that at  9 bars  but with a larger maximum wind speed of about 90 $\mps$.  The larger horizontal velocity divergence near the tropopause  causes more abundant turbulence and wave sources at this layer, and so stronger interactions of mean flow with turbulence and Rossby waves, generating stronger zonal flows.

Storms occur mostly near the equator, with almost no storms in mid-to-high latitudes. Notice that in our model setup the equilibrium potential temperature profile $\theta_{\rm{ref}}$  is independent of latitude, such that we can exclude any latitudinal dependent forcing as a possible cause of banded structure seen in our simulations.  Rather, any zonal banding, or latitude dependence of storms, must result  from the latitudinal  variation of $f$ and $\beta$ (where $\beta=d f/ d y$ is the gradient of Coriolis parameter with northward distance), which are the only sources that can introduce anisotropy in our simulations \citep{lian&showman2010}.

Diagnosing the dynamical mechanism for the latitudinal dependent storms is difficult. Instead we offer speculation based on the fact that $f$ and $\beta$ are the only possible sources of anisotropy (and therefore of any latitude dependence).  A possible reason is that the horizontal divergence of horizontal winds, $\nabla_p \cdot \mathbf{v}$, tends to be smaller at mid-to-high latitudes, implying smaller vertical velocities, which makes it more difficult to generate and maintain storms; the suppression of storms in turn further weakens vertical velocities by limiting horizontal temperature differences which are essential to drive horizontal divergence. There are two reasons that we expect small horizontal divergence at mid-to-high latitudes. First,  winds tend to be more geostrophic (the balance between Coriolis  and pressure gradient forces in the horizontal momentum equation [\ref{eq.momentum}]) in higher latitudes where the Coriolis parameter $f$ is larger, and this can lead to a smaller horizontal divergence. At low latitudes, winds have larger ageostrophic components, which results in a larger horizontal divergence. Thus moist instability can be more easily triggered. The importance of rotation can be characterized by the Rossby number, $Ro=U/\mathcal{L}f$, where $U$ and $\mathcal{L}$ are the characteristic horizontal wind speed and horizontal length scale, respectively. If $Ro \ll 1$, winds are nearly geostrophic. We can quantitatively estimate the latitude above which the flow tends to be geostrophic by taking $U \sim 100\mps $ and $\mathcal{L} \sim  10^6$ m, which are approximately the maximum relative velocity and the width of a local storm, respectively, and setting $Ro \sim 1$, and we have $\phi \sim 8^{\circ}$. This is qualitatively consistent with our simulations in which storms tend to clump inside $\pm 10^{\circ}$ latitudes (Figure \ref{f_bd15-1}).
Second, even if flow were geostrophic, the horizontal divergence is $\nabla_p \cdot \mathbf{v} \sim \frac{v}{a\tan\phi}$ where $v$ is meridional (north-south) velocity and $a$ is the radius of the brown dwarf. The divergence in pure geostrophic flow comes from the gradient of $f$ with respect to latitudes. It is easy to see that even in geostrophic flow, horizontal divergence becomes smaller in higher latitudes and larger in lower latitudes.
  However, the argument here does not mean that there is no vertical motions in high latitudes in general situations. In fact, if one imposes an independent meridional temperature difference to the atmosphere, it is easy to generate vertical motion and overturning circulations in high latitudes (e.g., \citealt{williams2003,lian2008}). The difficulty in generating high-latitude storms here is that the horizontal temperature differences that would be required for vertical motions are not independently generated but can only come from the existence of vertical motion (and the associated latent heating). This additional sensitivity allows for the suppression of storms in situations where vertical motions tend to be smaller, as at high latitudes.

\begin{figure}
\epsscale{1.}
\plotone{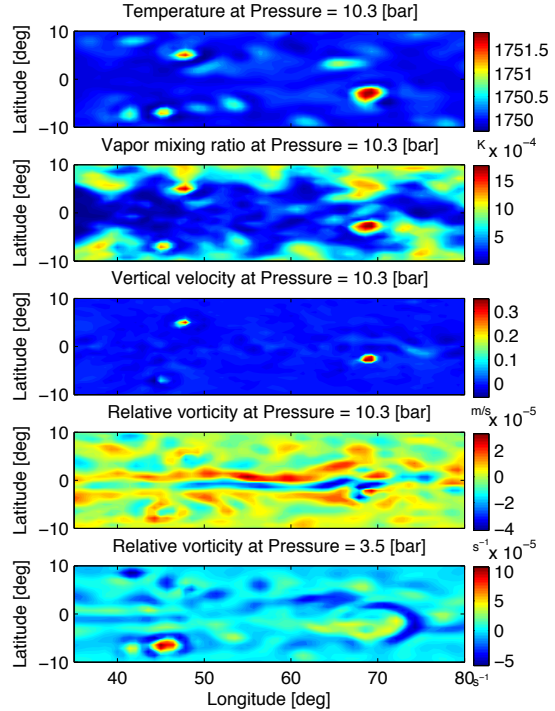}
\caption{Snapshots of temperature, vapor mass mixing ratio, vertical velocity and relative vorticity in a storm active region at about 1736 days for the nominal T dwarf simulation. Red in vertical velocity map represents upward direction. The condensation level for this simulation is at about 11 bars.}
\label{ttqv}
\end{figure}

Storms regions are buoyant, thereby causing ascending motion, which transports moist air upward from below, leading to condensation and latent heating ---  thereby maintaining the storms themselves. Therefore, storms are spatially well correlated with vapor mixing ratio and vertical velocities, as shown in the horizontal maps representing a local storm active area in Figure \ref{ttqv}. The vertical relative vorticity  $\zeta = \hat{k}\cdot \mathbf{\nabla}\times \mathbf{v}$ (lower two panels of Figure \ref{ttqv}) measures the local spin of  fluid in horizontal direction. If $\zeta$ has the same sign as the Coriolis parameter $f$, the storm is  cyclonic, whereas the storm is anticyclonic if $\zeta$ and $f$ have the opposite sign.   In the northern hemisphere, the base of storms (near 9 bars) generally have positive $\zeta$ and the top of storms (near 3.5 bar)  generally have negative $\zeta$. The dynamical picture for a single storm is that: because of the latent heating, the lower density of the storm column causes a greater vertical spacing of isobars, i.e., constant pressure lines bow  downward at the base of the storm and upward at the top of the storm. This causes a low-pressure center at the base of the storm and high pressure center at the top of the storm with respect to the surrounding environmental air at a given altitude. As a result, horizontal flow converges and diverges due to pressure gradient forces at the base and top of the storm, respectively. Meanwhile, the flow is accelerated by Coriolis force, which drives the flow to cyclonic at the base and anticyclonic at the top of the storms. 

\subsection{Zonal Jets}
\label{jets}
\begin{figure}      
\epsscale{1.}      
\plotone{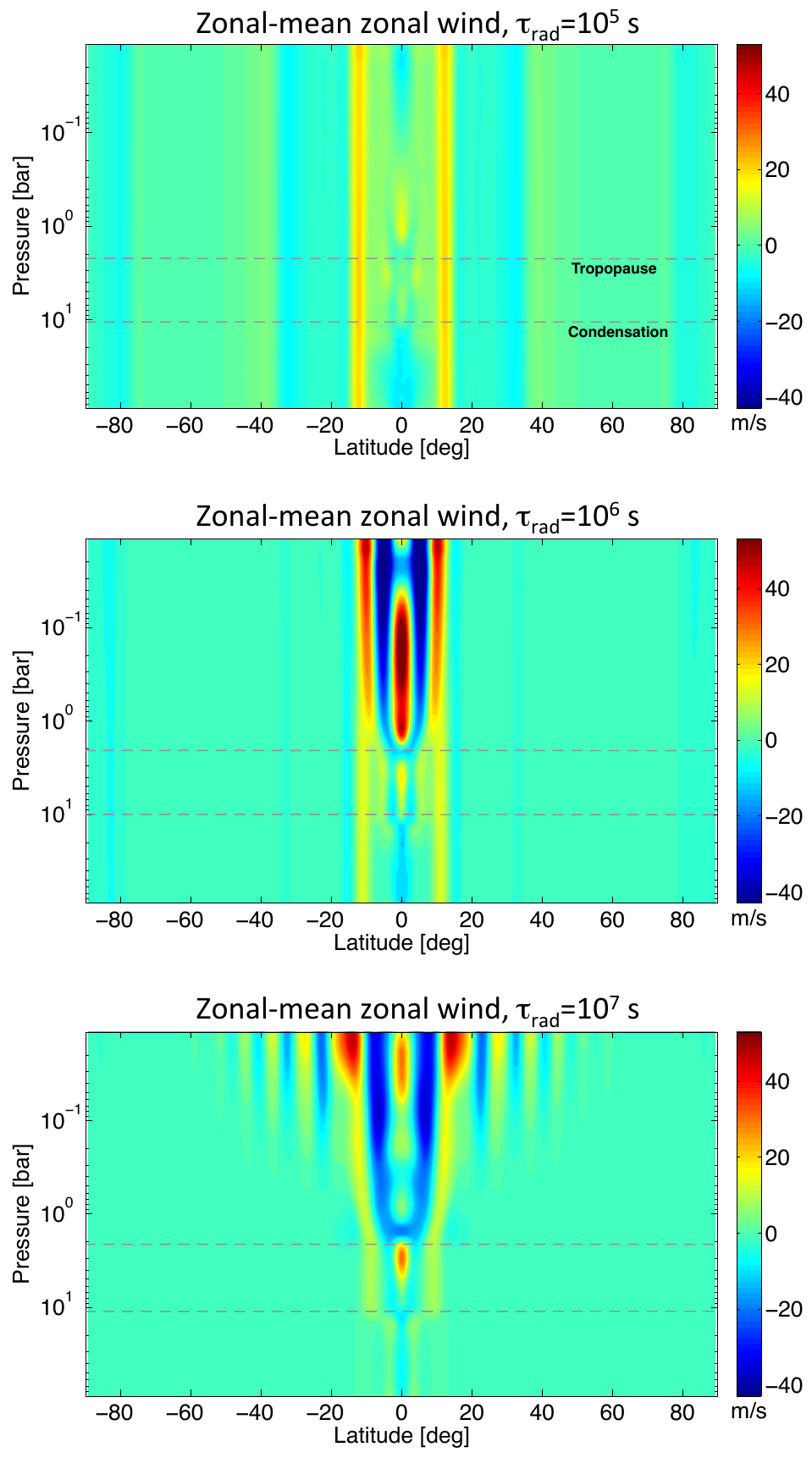}
\caption{Time-averaged zonal-mean zonal wind as a function of latitude and pressure for three simulations with different radiative timescale $\trad$: $10^5$ sec (upper), $10^6$ sec (middle) and $10^7$ sec (lower). Dashed lines in each panel mark the pressures for the condensation level and the tropopause, respectively. The tropopause here is defined as the vertical level above which the lapse rate $d\ln T/d\ln p$ is less than 0.268, which is slightly less than the adiabatic lapse rate $R/c_p = 0.2857$ but enough to stratify flows inside storms.  }
\label{f_bd-zonal}
\end{figure}

Large-scale latent heating drives global atmospheric circulation and forms zonal jets in our simulations, as discussed in Section \ref{latentheating}. The time-averaged zonal-mean zonal jet configuration from simulations with three different radiative timescales ($\trad=10^5, 10^6$ and $10^7$ sec, with other parameters the same as the typical T dwarf in Section \ref{tdwarf})  are shown in Figure \ref{f_bd-zonal}.   The results are averaged over about 1000 days after the models being equilibrated.
In general,  two strong eastward subtropical jets form at about $\pm 12^{\circ}$ latitudes and weak jets form in mid-to-high latitudes, which are symmetric about the equator.  At the equator, the equatorial jets are generally westward below the condensation level, and eastward equatorial jets appear just above the condensation level. The equatorial jet speed increases with height to the tropopause because of the  baroclinic structure by latent heating,  with its strength depending on radiative timescale. Here, baroclinic means that constant density surfaces are \emph{not} aligned with constant pressure surfaces, whereas barotropic means that the two surfaces are aligned. The local maximum jet speed near the tropopause is caused by the strong dynamical perturbations from eddies generated at the tops of storms. Jets below the condensation level are generally weak,  and the subtropical jets are presumably driven by the Coriolis force\footnote{Notice that because of the fast rotation, the Coriolis parameter $f$ has a large magnitude of about $10^{-4} ~\rm{s^{-1}}$  even at $10^{\circ}$ latitude.}  on the meridional circulation  in the deep atmosphere that results from the circulation of the upper active layer \citep{haynes1991,showman2006,lian2008}. 
The jets extend into the upper stably stratified atmosphere. The circulation above the tropopause probably emerges from the absorption, dissipation and breaking of upward propagating waves that are generated at the tropopause \citep{showman&kaspi2013}. There have been extensive studies showing that the mechanical, wave-induced forcing is the dominant driver for stratospheric circulation on Earth despite the existence of equator-to-pole thermal forcing (see review by, for example, \citealp{andrews1987, haynes2005}). In conditions of our simulated atmospheres where isotropic equilibrium thermal structure is imposed, the wave-induced mechanical forcing should be responsible for the stratospheric circulation. In fact, we have observed upward propagating waves from levels perturbed by latent heating in our simulations, which supports our hypothesis. 
The jet structure exhibits differences with different radiative timescale $\trad$, as $\trad$ can affect the rate at which the characteristic horizontal temperature differences and dynamical perturbations are damped.  First, the short-$\trad$ model shows nearly barotropic jet structure, whereas relative high $\trad$ models show baroclinic structure. Second, the jet speed is generally larger for the relatively large-$\trad$ model, presumably because there is more time for jets to pump up before dynamical perturbations are damped out; this relation has been formulated in \cite{showman&kaspi2013} using the quasi-geostrophic theory. The jet structure  within about 2 -- 3 pressure scale heights of the upper boundary for the model with $\trad=10^7$ sec (the lower panel in Figure \ref{f_bd-zonal}, jets from $20^{\circ}$ to $50^{\circ}$ latitude) is likely affected by the upper boundary conditions. We have tested models with higher upper boundaries ($10^{-3}$ bar and less), and the dynamics deeper than about 3 pressure scale heights from the upper boundary remains almost the same as in the original model. We conclude that despite the imperfection of numerics near the upper boundary for model with $\trad=10^7$ sec, the dynamics presented here for the atmosphere below about 0.1 bar is physical.  

\subsection{Area Fraction of Storms}
\label{sec-fraction}
\begin{figure}
\epsscale{1.}
\plotone{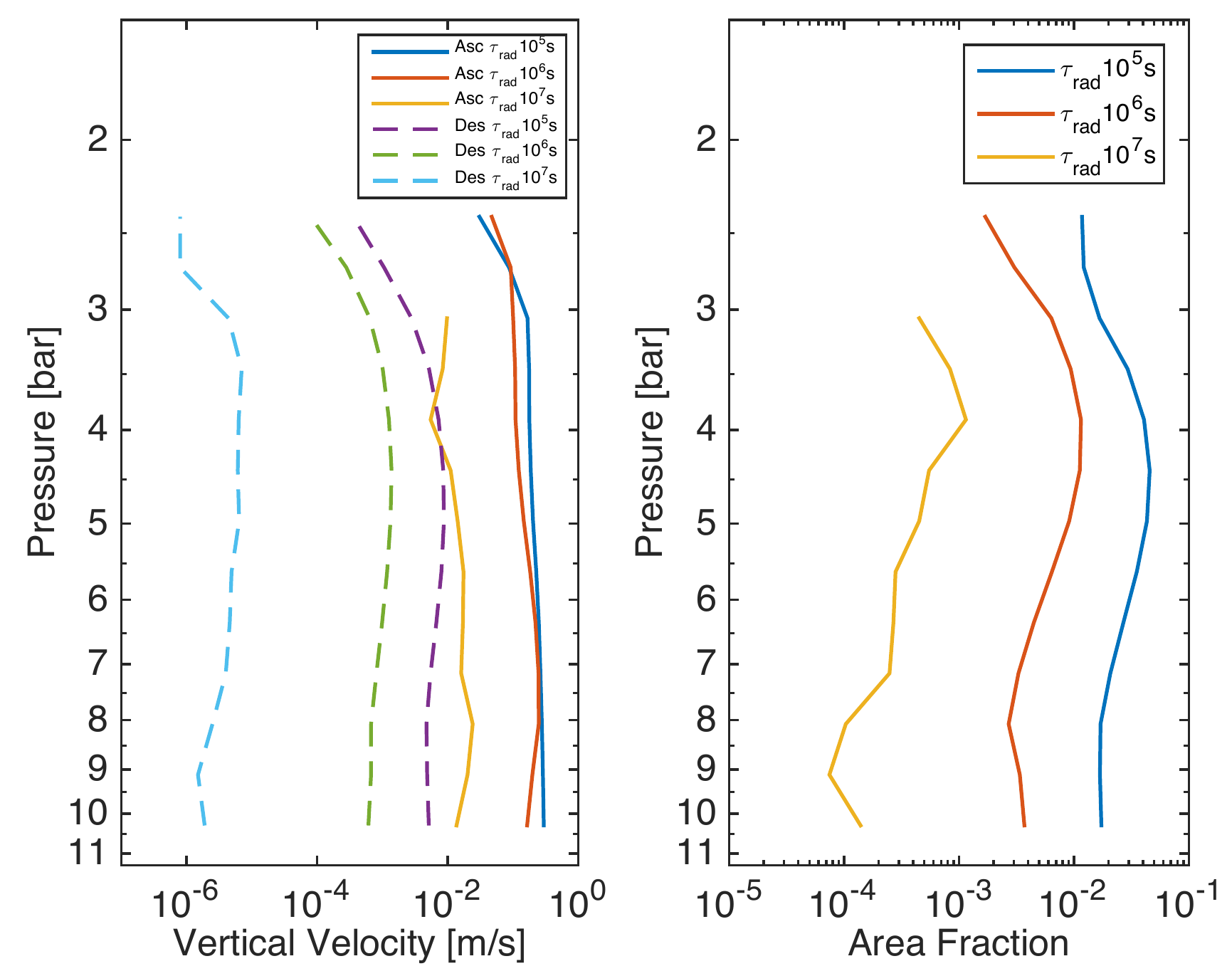}
\caption{\emph{Left}: The absolute values for time-averaged ascending vertical velocity ($\mps$) inside storms (solid lines) and descending velocity outside storms (dashed lines) as a function of pressure for the models with $\trad = 10^5,10^6$ and $10^7$ sec. \emph{Right}: Area fraction of storms $\sigma_s(p)$ as a function of pressure for the same models.  }
\label{wsigma}
\end{figure}

The small area fraction of moist plumes has been visually shown in Figure \ref{f_bd15-1} and \ref{ttqv} for a typical T dwarf model, in which the discretized warm areas only occupy a small fraction of the area in low latitudes where storms are active. Not only the area occupied by storms is small, the sizes of individual storms are also small, having diameter of around $2^{\circ}$ (about the length of 1700 km, or three grid cells) above the condensation level at around 10 bars for typical storms in all our models. As shown in Figure \ref{f_bd15-1}, the storms slightly expand near the tropopause where flows experience stratification and expand laterally. Here we display a more quantitative measurement of the storm fraction in models with different radiative timescale $\trad$. The storms  are defined roughly between the  condensation level and the tropopause, and they should have both a saturated mixing ratio of vapor ($q\geq q_s$) and an upwelling velocity. Using the upward vertical velocity as an indicator for storms is reasonable in our case because, as will be shown below, the upward vertical velocities inside storms are much larger than the descending velocities outside storms. We tested the sensitivity of this criteria by choosing different numbers, for example, $q\geq 0.98q_s$ or $q\geq 1.02q_s$, and these different criteria do not affect the results. We define regions satisfying these two criteria as being inside storms. Regions not satisfying these criteria are defined as regions outside storms. We only count areas within about $\pm 9^{\circ}$ latitudes since this is the primary region where storms occur.  We first count vertical velocities as a function of pressure inside and outside storms using instantaneous snapshots of vertical velocity field from simulations, then define the area fraction of storms as\footnote{This is a definition based on continuity argument, consistent with that defined in Section \ref{mechanism}.} $\sigma_s(p) = |\omega_d(p)/ \omega_a(p)|$, where $\omega_d(p)$ and $\omega_a(p)$ are the  spatially averaged vertical velocity outside storms  and  inside storms, respectively. Finally, $\sigma_s(p)$ is averaged over many snapshots at different simulation time over about 1000 days after the simulation equilibrates.
The results are shown in Figure \ref{wsigma} as a function of pressure for models with $\trad=10^5, 10^6$ and $10^7$ sec. In the left panel, the vertical velocity is in a modified log-pressure coordinate $-Hd(\ln p)/dt$, where $H$ is pressure scale height; this velocity is approximately equal to the vertical velocity in height coordinates. This is a standard way of representing vertical velocity in pressure coordinates (e.g., \citealp{andrews1987}). Physically, the quantity $d(\ln p)/dt$ is the vertical velocity expressed in units of scale heights per second, with positive being downward.  Multiplying by $-H$ converts this to the vertical velocity in $\rm{m~s^{-1}}$, with positive being upward.  As long as the structure of isobars does not change rapidly with respect to $z$ over time, this quantity will be approximately equal to the vertical velocity in height coordinates.  The magnitude of descending vertical velocities clearly decrease by order of magnitude with increasing $\trad$.  The ascending velocities are similar for $\trad=10^5$ and $\trad=10^6$ sec, but they are a factor of $\sim 5$ smaller for $\trad=10^7$ sec. As a result, the area fraction of storms decreases orders of magnitude as $\trad$ increases. We quantitatively explore the mechanism controlling the area fraction in Section \ref{scalling}.

\subsection{Enhanced Abundance of Condensate}
\begin{figure}
\epsscale{1.}
\plotone{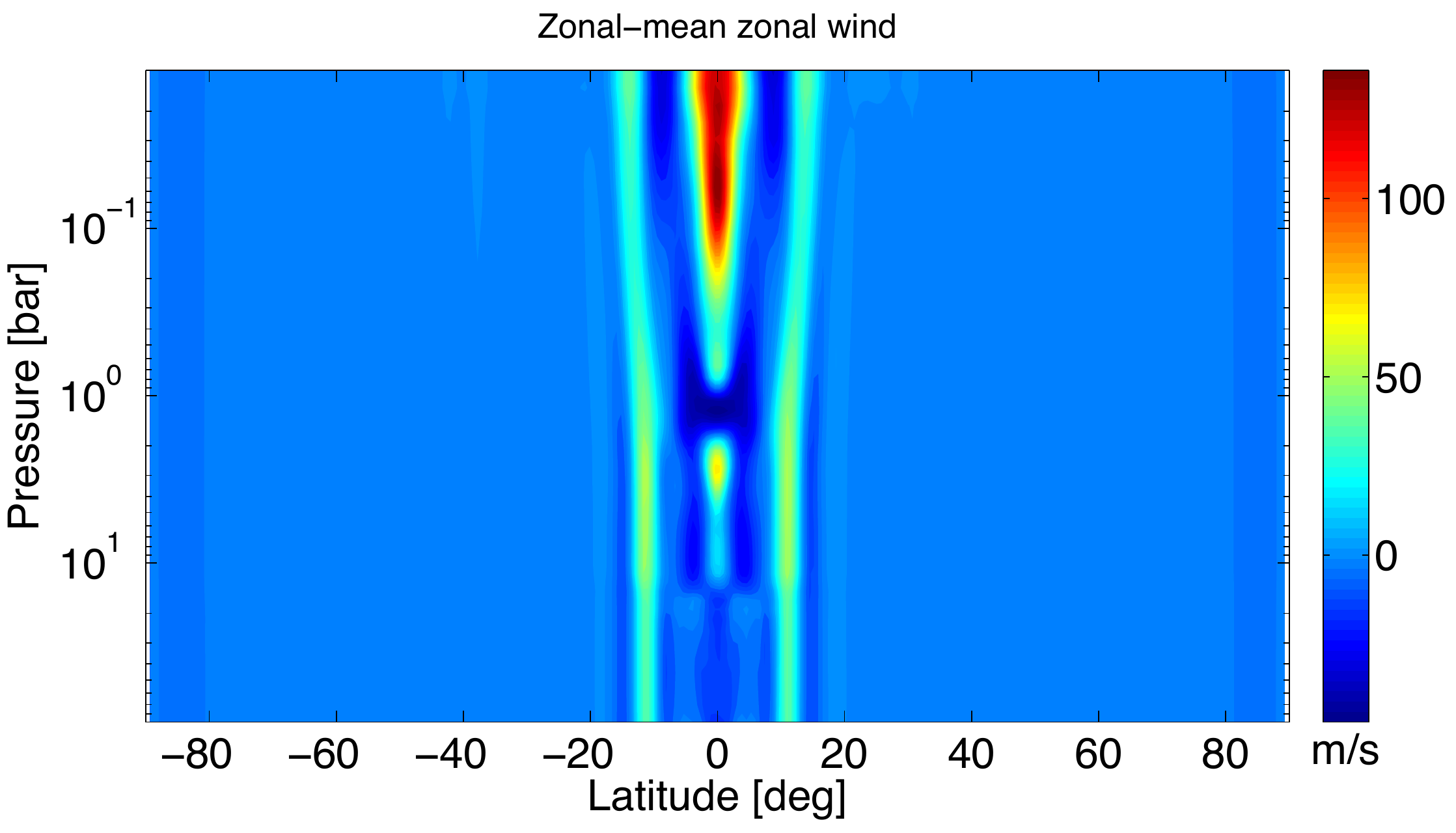}
\caption{Time-averaged zonal-mean zonal wind for a T dwarf model with 3 times solar abundance and $\trad=10^6$ sec. }
\label{m3t6}
\end{figure}

Giant planets tend to have metal-rich atmospheres, having condensed out of the gas-depleted disks around preferentially metal-enriched host stars \citep{gonzalez1997}. In the context of our model, an enhanced metallicity means a greater abundance of silicate vapor, a higher latent heating and thus a stronger atmospheric circulation. We have ran models with  three times solar abundance of silicate vapor, representing  the possible conditions of  directly imaged EGPs. Generally, the  basic pattern of the condensation cycle and the zonal jet configuration are similar to the solar abundance models, but with larger temperature perturbations  (proportional to the abundance of vapor), active storms occurring up to slightly higher latitudes and larger wind speeds. Figure \ref{m3t6} shows the time-averaged zonal-mean zonal wind from a model with three times solar abundance (typical for heavy element abundances on Jupiter) and $\trad=10^6$ sec. The zonal jet structure is very similar to that of our model with solar abundance (middle panel of Figure \ref{f_bd-zonal}), except that the winds are enhanced by a factor of several.

\section{DISCUSSION and conclusions}
\label{discussion}

\subsection{What Controls the Area Fraction of Moist Plumes?}
\label{scalling}

Here we construct a simple model to quantitatively understand the area fraction of storms shown in Section \ref{sec-fraction} by constructing scaling relations for the governing equations (\ref{eq.momentum}) -- (\ref{thermal}). The physical picture of the model comprises statistically steady  storms and subsidence outside the storms. At its top, the storm center has high pressure relative to the surroundings which can drive an outward divergent flow; then the high-entropy air radiatively cools over time during the slow subsidence, reaching almost the same temperature as the environment at the condensation level by requirement of steady state. Here we ignore the density variations due to rainout of condensate.  The area fraction of  storms $\sigma_s$ is given by requirement of continuity
\begin{equation}
\sigma_s \sim |\frac{\omega_d}{\omega_a}|,
\label{eq.0}
\end{equation}
where $|\omega_d| \ll |\omega_a|$.  As presented in Section \ref{tdwarf}, storms mostly occur at low latitudes where the Rossby number is large ($\gtrsim 1$). Near the top of the storm, the horizontal force balance is primarily between advection and pressure gradient force. In low latitudes, the Coriolis force could still have a nontrivial magnitude compared to the advection force. Including the Coriolis force in our scaling induces only a mild correction to our final result  (Equation [\ref{fraction}]), but does not change our conclusion in this section. Therefore for the sake of a clearer illustration of the physical mechanism controlling the fractional area of storms, using the force balance between advection and pressure gradient is reasonable. Therefore the  balance in horizontal momentum equation (\ref{eq.momentum}) is  $\mathbf{v}\cdot\nabla_p \mathbf{v} \sim -\nabla_p \Phi$,  which to order of magnitude reads 
\begin{equation}
\frac{U^2}{\mathcal{L}} \sim  \frac{\Delta\Phi}{\mathcal{L}},
\label{eq.1}
\end{equation}
 where  $\Delta \Phi$ is the horizontal difference in gravitational potential between the top of the storm and its surroundings on a constant pressure surface.
 From hydrostatic equilibrium (Equation \ref{eq.hydro}), we can estimate the pressure difference inside and outside the storm by integrating over the column:
 \begin{equation}
 \Delta \Phi \sim R \delta \ln p \Delta T,
 \label{eq.2}
 \end{equation}
 where $\delta \ln p$ is the vertical difference in log-pressure from the bottom to the top of the storm and $\Delta T$ is the characteristic horizontal temperature difference inside and outside the storm.
 From the continuity Equation (\ref{eq.cont}), the horizontal divergence at the storm given by  $\nabla_p \cdot \mathbf{v}\sim U/\mathcal{L}$, is balanced by vertical divergence of ascent inside the storms, This implies
 \begin{equation}
 \frac{U}{\mathcal{L}} \sim \frac{\omega_a}{\delta p},
 \label{eq.3}
 \end{equation}
 where $\delta p$ is the difference in pressure from the bottom to the top of the storm.
  Combining Equation (\ref{eq.1}), (\ref{eq.2}) and (\ref{eq.3}), and assuming constant vertical velocity, the ascending velocity $\omega_a$ can be estimated by
 \begin{equation}
 \omega_a \sim \frac{\delta p \sqrt{R\Delta T \delta \ln p}}{\mathcal{L}}.
 \end{equation}
 To estimate the descending velocity, we use the thermodynamic energy equation (\ref{thermal}), and assume that  vertical advection of the potential temperature is much larger than the horizontal advection. This is reasonable near the tropopause where vertical difference of potential temperature is much larger than the horizontal differences. We then can obtain the balance between radiative cooling and vertical advection, which to order of magnitude reads:  $\omega_d \frac{\delta \theta}{\delta p} \sim \frac{\theta-\theta_{\rm{ref}}}{\trad}$, where $\delta \theta$ is the vertical difference in potential temperature outside storms between pressure levels corresponding to the bottom and the top of storms. Imagining a thermodynamics loop where air rises in storms and subsides in between storms, we expect that at the altitude of the storm top, the environmental air outside storms has just been detrained from the top of the storm, and therefore that the potential temperature of the storm air and environmental air are the same at the pressure of the storm top. Likewise, in a closed thermodynamic loop we expect that the potential temperature of environmental  and storm air are equal at the storm bottom. To close the system, we assume that in a global-mean and steady state, the higher-entropy air descending from the top of storms radiates away most of its entropy gained from latent heating, and relaxes to nearly the reference temperature when the air reaches the bottom of storms, which implies  $\delta \theta \sim \theta-\theta_{\rm{ref}}$. Assuming constant $\omega_d$, we can estimate the descending velocity as 
 \begin{equation}
 \omega_d \sim \frac{\delta p}{\trad}.
 \end{equation}
 This equation simply states that the rate of descent is bottlenecked by the efficiency of radiation: in order for the air outside storms to descend over the vertical height of a storm, the air has to lose entropy (since the environment is stratified), and thus this descent must occur on timescales comparable to the radiative time constant. 
 Finally, the area fraction can be obtained by substituting $\omega_d$ and $\omega_a$ into Equation (\ref{eq.0}):
 \begin{equation}
 \sigma_s \sim \frac{\mathcal{L}}{\trad \sqrt{R\Delta T \delta \ln p}}.
 \label{fraction}
 \end{equation}
 This is essentially a timescale comparison, where $ \mathcal{L}/\sqrt{R\Delta T \delta \ln p}$ is the dynamical ascent timescale driven by CAPE and $\trad$ is the timescale driven by radiative dissipation.
 According to results in Section \ref{results}, taking $\mathcal{L}\sim 10^6$ m,  $\Delta T \sim 2.5$ K, $\delta \ln p \sim 1.3$ and $\trad \sim 10^5, 10^6$ and $10^7$ sec, we have area fraction $\sigma_s \sim 10^{-1}, 10^{-2}$ and $10^{-3}$, respectively. Compared to results from simulations in the right panel of Figure \ref{wsigma}, our analytical model to order of magnitude agrees well with the maximum area fraction for different $\trad$. The area fraction from our numerical results show variation as a function of pressure which in general  are off by a factor of a few to ten compared to our analytical model.  Given the simplicity of the scaling theory, we can explain the order of magnitude decrement of area fraction with increasing $\trad$, illustrating the important regulation of radiation on the moist convection.

Real cloud formation exhibits many complexities not accounted for in this simple scaling theory. For example, the intertropical convergence zone in Earth's tropical region shows organized regions of vigorous cumulus convection, containing transient cloud clusters rather than simply regions of steady-state precipitation and mean updrafts. This is a result of interactions between local cumulus convection and large-scale atmospheric circulation (\citealp{holton2012}, Chapter 11). The large-scale latent heating scheme in our model does not represent the small-scale cumulus convection, but rather the hydrostatic interaction of the storms with their surroundings. To understand the interactions between sub-grid moist convection and large-scale flow, we need a better parameterization of moist convection in future studies.   Also, radiative feedback by cloud particles can  play an important role in the development of cumulus clouds.  Our analysis here will be tested using more realistic models in future studies.

 \subsection{Thermal Structure}
\begin{figure}
\epsscale{1.}
\plotone{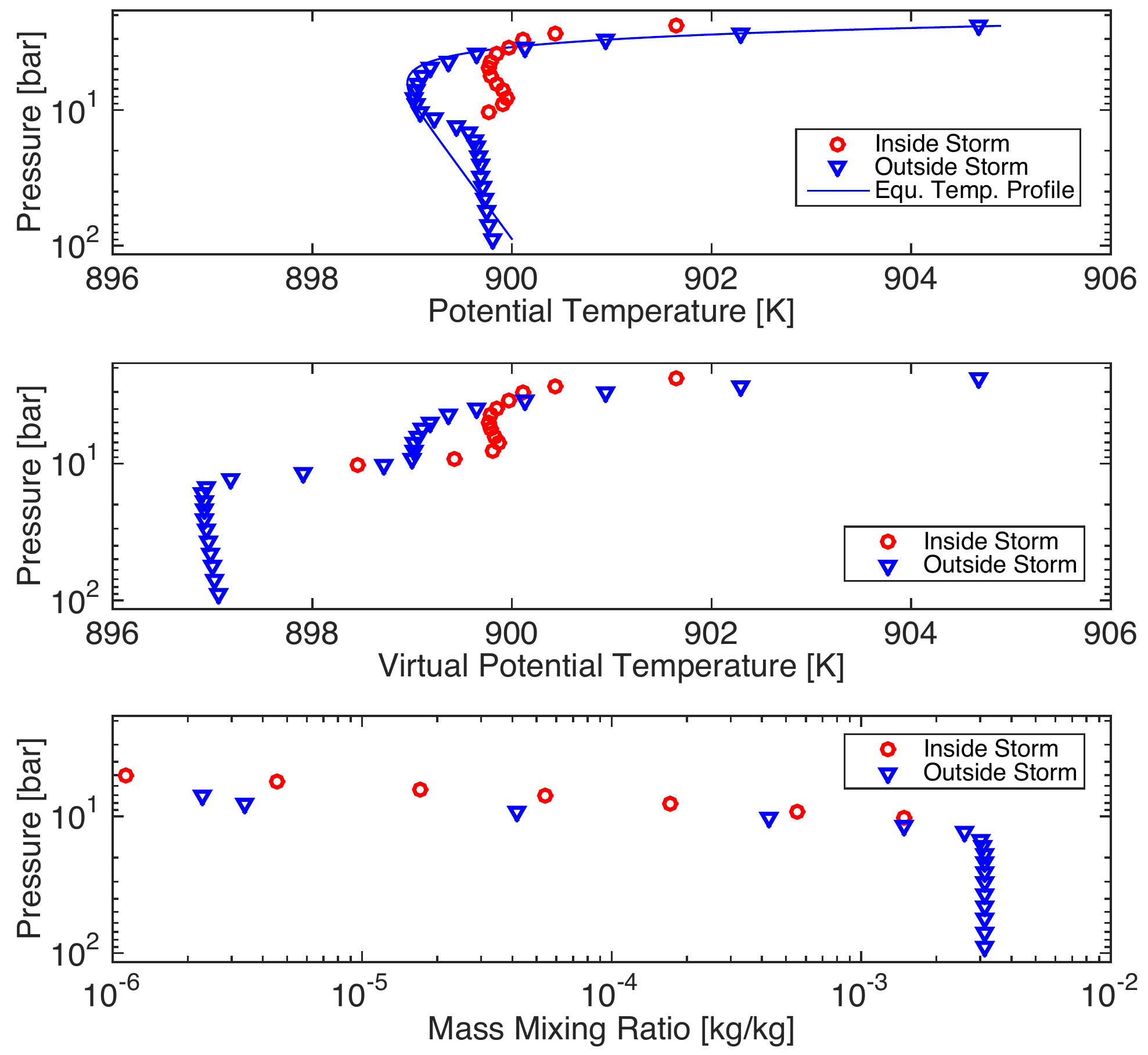}
\caption{Thermal structure of the nominal simulation shown in Section \ref{results} with radiative timescale $\trad = 10^6$ sec. The upper panel shows potential temperature $\theta$ for air inside and outside storms as indicated by the labels, as well as the equilibrium temperature profile. The middle panel shows the corresponding virtual potential temperature $\theta_v$ profile, and the lower panel shows the mixing ratio of condensable vapor for air outside and inside storms. }
\label{thermalstructure}
\end{figure}

The thermal structure of the atmosphere can be affected directly by latent heating via its effect on temperature, and indirectly by the molecular weight effect via introducing a stratification layer above the condensation level. The upper panel of Figure \ref{thermalstructure} shows potential temperature $\theta$ as a function of pressure for air outside and inside of storms from the nominal simulation with radiative timescale $\trad = 10^6$ sec in Section \ref{results}, and the middle panel shows the corresponding virtual potential temperature $\theta_v$ profile. The virtual potential temperature $\theta_v$ is defined as $\theta_v=(\frac{1+q/\epsilon}{1+q})\theta$, where $\epsilon = m_v/m_d$ is the ratio of molecular mass of condensable species and the dry air, and $q$ is mass mixing ratio. It can be viewed as the theoretical potential temperature that a dry air parcel would have if the dry parcel has the same pressure and density as the moist air, so $\theta_v$ is a direct measurement of density. The solid line in the upper panel is the equilibrium background temperature profile prescribed by Equation (\ref{eq.theta}). In the deep convective region, temperatures do not exactly follow the reference profile because dry motions tend to neutralize the thermal structure by having nearly constant $\theta$. 
However, due to rainout of condensates as shown in the lower panel of Figure \ref{thermalstructure},  the layer just above the condensation level is stratified, and is stable against dry convection. As a result, this thin layer just above the condensation level is not neutralized by dry motions. The stratification of this thin layer is better illustrated by looking at the virtual potential temperature $\theta_v$ profile in the middle panel, in which $\theta_v$ increases  with height despite the fact that $\theta$ actually decreases with height. Note that increasing $\theta_v$ with height implies stratification against dry convection, accounting for both temperature and molecular weight gradients. The red circles represent (virtual) potential temperatures inside storms, and their profile is nearly close to a moist adiabat. Interestingly, at the bottom of the stratosphere, ``overshooting'' of moist plumes occurs, in which temperature inside storms (red circles) has lower temperature than surroundings. This suggests that ascending moist plumes penetrate into the stratosphere, inducing horizontal flow divergence described in Section \ref{tdwarf}.
Due to the dry subsidence, layers immediately above  the condensation level outside storms are unsaturated, which leads to decreasing mean molecular weight  with decreasing pressure. And this contributes to a strong stratified background environmental profile at pressures near the condensation level. Interestingly, it also results in a lower density of the background air than that of the moist upwelling plumes. It can be seen in the middle panel of Figure \ref{thermalstructure}, where air inside storms (red circle) has a  lower $\theta_v$ than background air (blue triangle) right above the condensation level. Overall, the molecular weight effect is likely important in regulating the way storms occur and the way they interact with their environment.  This is similar to regional moist convection simulations in atmospheres of giant planets where water has larger molecular weight than hydrogen-helium mixture \citep{nakajima2000,sugiyama2014,Li2015}.  Indeed, simulations from models without molecular weight effect have more vigorous storm activities than those with molecular weight effect. However, how this regulation would work in a hydrostatic manner is unclear, and needs more diagnosis in future work. 

 \subsection{Implication for Observations}
Due to the lack of radiative transfer and cloud particles, we are unable to directly compare the simulated  variability to the observed near-IR variability. Still, our results have important implications for observations. Storms driven by moist instabilities can extend vertically over several pressure scale heights, reaching the photosphere.  The vertical velocity inside the large-scale hydrostatic storms is high, and condensed particles can be lofted up to the storm top, forming cumulus clouds and  inducing IR brightness variability.  The storms can evolve on timescales of hours to days, and the cumulus clouds would be patchy due to the spatially inhomogeneous moist convection, and thus can help to explain patchy clouds inferred in the rapid evolving near-IR light curves and observationally inferred surface maps of brown dwarfs.  Recently, \citet{karalidi2015} and \citet{karalidi2016} present retrieval surface temperature maps for a few brown dwarfs based on   near-IR light curves. Interestingly, the deduced temperature anomaly patterns are much larger than the expected Rossby deformation radius ($\sim 10^7$ m). It may be caused by a cluster of  storms over a large fraction of the globe similar to that shown in Figure \ref{ttqv},  which may produce a broad envelope of patchy clouds that  represent as a single large spot.

The L/T transition occurs over a narrow range of effective temperature accompanied with a \emph{J}-band brightening (e.g., \citealp{allard2001,burrows2006,saumon2008}), and its details remain poorly understood. Hypotheses include a change of sedimentation efficiency for condensates \citep{knapp2004} or that the cloud deck gradually becomes patchy as clouds form progressively deeper with increasing spectral type, allowing contributions from greater flux emitted from deeper levels \citep{ackerman2001, burgasser2002, marley2010}. However, the detailed mechanisms for cloud breaking during the L/T transition are yet unclear.  Here we propose that,  the area fraction of moist convection can help to support the idea of cloud breaking during the L/T transition.  Moist convection occurs when large CAPE is available, that is, the condensation level should be much lower than the tropopause. This can also be quantified from Equation (\ref{fraction}) that  relatively large $\delta \ln p$ and long radiative time constant $\trad$ are needed to produce a small storm fraction. In the hotter L dwarfs, clouds first condense close to upper stratified atmosphere (e.g., \citealp{tsuji2002, burrows2006}), so moist convection can not happen, and cloud morphology may be dominated by the  stratus clouds formed by more gradual processes such as transport by waves \citep{freytag2010} or large-scale atmospheric flow \citep{showman&kaspi2013}. For the cooler dwarfs near the L/T transition, the condensation level gradually sinks below the tropopause, moist convection thus can occur with increasing CAPE, producing patchy cumulus clouds.
 Also, as condensation level moves to a larger pressure, the radiative timescale $\trad$, which can be approximated by $\trad \sim \frac{p}{g}\frac{c_p}{4\sigma T^3}$ where $\sigma$ is the Stefan-Boltzmann constant, may become larger. According to our results, the larger $\trad$ can also decrease the storm area fraction.  The changing of cloud patchiness during the L/T transition can be a natural consequence of the change of CAPE and radiative timescale with increasing spectral type. We predict that, if latent heating dominates cloud formation processes in atmospheres of BDs and directly imaged EGPs, the fractional coverage area of clouds gets smaller as the spectral type goes through the L/T transition from high to lower effective temperature.   Future more realistic models are needed to test our hypothesis.   


\subsection{Summary}
Latent heating from condensation of various chemical species in brown dwarf atmospheres is important for shaping the atmospheric circulation and influencing  cloud patchiness. We illustrated the dynamical mechanisms of latent heating using an idealized atmospheric circulation model that includes a condensation cycle of silicate vapor with the molecular weight effect included. For typical T dwarf models, zonal jets can be driven by large-scale latent heating. Temperature maps show inhomogeneous storm patterns, which evolve on timescales of hours to days and can extend vertically over a pressure scale height  or more to the tropopause. The fractional area of the brown dwarf covered by active storms is small. Based on a simple analytic model, we quantitatively explain the fractional area of storms, and predict its dependence on radiative timescale and convective available potential energy. 
 Our results have important implications for the observed near-IR variability and the cloud properties across the L/T transition. Further general circulation models with realistic clouds and radiative transfer are needed for better investigation of the global circulation.

\parskip = \baselineskip
We thank Xi Zhang for helpful discussion. This work was supported by NASA Headquarters under the NASA Earth and Space Science Fellowship Program and NSF grant AST 1313444 to APS.

\bibliographystyle{apj}
\bibliography{draft}

\end{document}